\documentclass[12pt]{article}
\usepackage[left=2cm, right=2cm, bottom=2cm ,top = 2cm, footskip=1cm]{geometry}
\usepackage[font=small,labelfont=bf]{caption}
\usepackage{setspace}
\usepackage{lineno}

\usepackage{chngcntr}

\usepackage[dvipsnames]{xcolor}

\usepackage{hyperref}

\usepackage[numbers,sort&compress]{natbib}
\usepackage{graphicx}
\usepackage{amsmath}
\usepackage{upgreek}

\usepackage[section]{placeins}

\title{Vagus nerve stimulation: Laying the groundwork for predictive network-based computer models}
\author{John F. Ingham$^{1*\P}$ \and Frances Hutchings$^{1\P}$\and Paolo Zuliani$^{4}$ \and Yujiang Wang$^{1,2,3}$ \and Sadegh Soudjani$^{5}$ \and Peter N. Taylor$^{1,2,3}$}
\begin{document}

\maketitle

\begin{enumerate}
\item{CNNP Lab (www.cnnp-lab.com), Interdisciplinary Computing and Complex BioSystems Group, School of Computing, Newcastle University, Newcastle upon Tyne, United Kingdom}
\item{Faculty of Medical Sciences, Newcastle University, Newcastle upon Tyne, United Kingdom}
\item{UCL Queen Square Institute of Neurology, Queen Square, London, United Kingdom}
\item{Dipartimento di Informatica, Università di Roma ``La Sapienza'',
Rome, Italy}
\item{Max Planck Institute for Software Systems, Kaiserslautern, Germany}
\end{enumerate}

\begin{center}
* Corresponding author\\
j.f.ingham2@newcastle.ac.uk \\  
$^{\P}$ These authors contributed equally to this work
\end{center}

\section{Abstract}
Vagus Nerve Stimulation (VNS) is an established palliative treatment for drug resistant epilepsy. While effective for many patients, its mechanism of action is incompletely understood. Predicting individuals’ response, or optimum stimulation parameters, is challenging. Computational modelling has informed other problems in epilepsy but, to our knowledge, has not been applied to VNS. 

We started with an established, four-population neural mass model (NMM), capable of reproducing the seizure-like dynamics of a thalamocortical circuit. We extended this to include 18 further neural populations, representing nine other brain regions relevant to VNS, with connectivity based on existing literature. We modelled stimulated afferent vagal fibres as projecting to the nucleus tractus solitarius (NTS), which receives input from the vagus nerve \textit{in vivo}.

Bifurcation analysis of a deterministic version of the model showed higher background NTS input made the model monostable at a fixed point (FP), representing normal activity, while lower inputs produce bistability between the FP and a limit cycle (LC), representing the seizure state.

Adding noise produced transitions between seizure and normal states. This stochastic model spent decreasing time in the seizure state with increasing background NTS input, until seizures were abolished, consistent with the deterministic model. 

Simulated VNS stimulation, modelled as a 30 Hz square wave, was summed with the background input to the NTS and was found to reduce total seizure duration in a dose-dependent manner, similar to expectations \textit{in vivo}.

We have successfully produced an \textit{in silico} model of VNS in epilepsy, capturing behaviour seen \textit{in vivo}. This may aid understanding therapeutic mechanisms of VNS in epilepsy and provides a starting point to (i) determine which patients might respond best to VNS, and (ii) optimise individuals’ treatments.

\section{Introduction}
Epilepsy is a common neurological condition, both disabling and potentially dangerous, with an estimated lifetime prevalence of 0.76\% worldwide~\cite{fiestPrevalenceIncidenceEpilepsy2017a}. First line treatment is pharmacological, but this is not completely effective in up to a third of cases~\cite{sultanaIncidencePrevalenceDrugResistant2021} and also prone to common side effects~\cite{elliottEfficacyVagusNerve2011}.

Vagus Nerve Stimulation (VNS) is among the treatments used for drug resistant cases. VNS can be used for multiple types of epilepsies, and is generally regarded as palliative, reducing seizure frequency and duration by approximately 50\% in around half of patients~\citep{brucknerCognitivePsychologicalSide2021, degiorgioProspectiveLongTermStudy2000}. It is safe for long term use, and benefits can sometimes improve for as long as two years after starting therapy~\citep{Morris1999}.  It is licensed, not only for epilepsy, but also for the treatment of refractory depression. Other suggested use includes conditions as diverse as heart failure, migraine, cluster headache, post-traumatic stress disorder, rheumatoid arthritis, fibromyalgia, obesity and inflammatory bowel disease~\citep{johnsonReviewVagusNerve2018}. The vagi are a pair of cranial nerves, originating in brainstem nuclei, which innervate structures in the head, neck, thorax and abdomen. Both efferent (voluntary branchial motor and involuntary visceral motor) and afferent (general somatosensory, general and special visceral sensory) fibres are present.

In the context of epilepsy treatment, VNS is applied by a device implanted subcutaneously in the chest, connected to electrodes on the left vagus nerve as it runs through the neck, aiming to preferentially stimulate afferent fibres. Through a combination of clinical experience and engineering expediency, practice has settled on a 30Hz monophasic square-wave signal, with a pulse width of 250-500$\mu$s at a current between 0.25mA and 3.5mA. This is then applied cyclically for 30s in every 300s. To minimize side effects, initial settings are conservative, and parameters are adjusted during clinic visits, based on clinical experience and trial and error~\citep{thompsonReviewParameterSettings2021, barthSexHormonesAffect2015}. 

Despite its use in epilepsy, and ongoing clinical and animal research, the mechanism of VNS in epilepsy remains incompletely understood. Furthermore, predicting which patients will respond to VNS is imprecise, with few biomarkers identified which correlate to any significant degree~\citep{workewychBiomarkersSeizureResponse2020}. 

Computational modelling is a technique to capture known aspects of brain function with systems of differential equations. It has provided insights into epilepsy surgery prognosis~\cite{hutchingsPredictingSurgeryTargets2015,sinhaSilicoApproachPresurgical2014}, identifying factors predicting successful seizure termination by cortical stimulation~\cite{taylorComputationalStudyStimulus2014}, and predicting optimal methods for disrupting seizures with optogenetic stimulation ~\cite{Zaaimi2020}. To our knowledge, no computational model of VNS in epilepsy currently exists. 

Here we present our newly developed computational model of VNS in epilepsy and demonstrate that it is able to reproduce behaviours found clinically. We envisage this model will serve as a platform for future investigations into VNS in epilepsy.

\section{Materials and methods}

\begin{figure}
	\centering
	\includegraphics[width= 15.4cm]{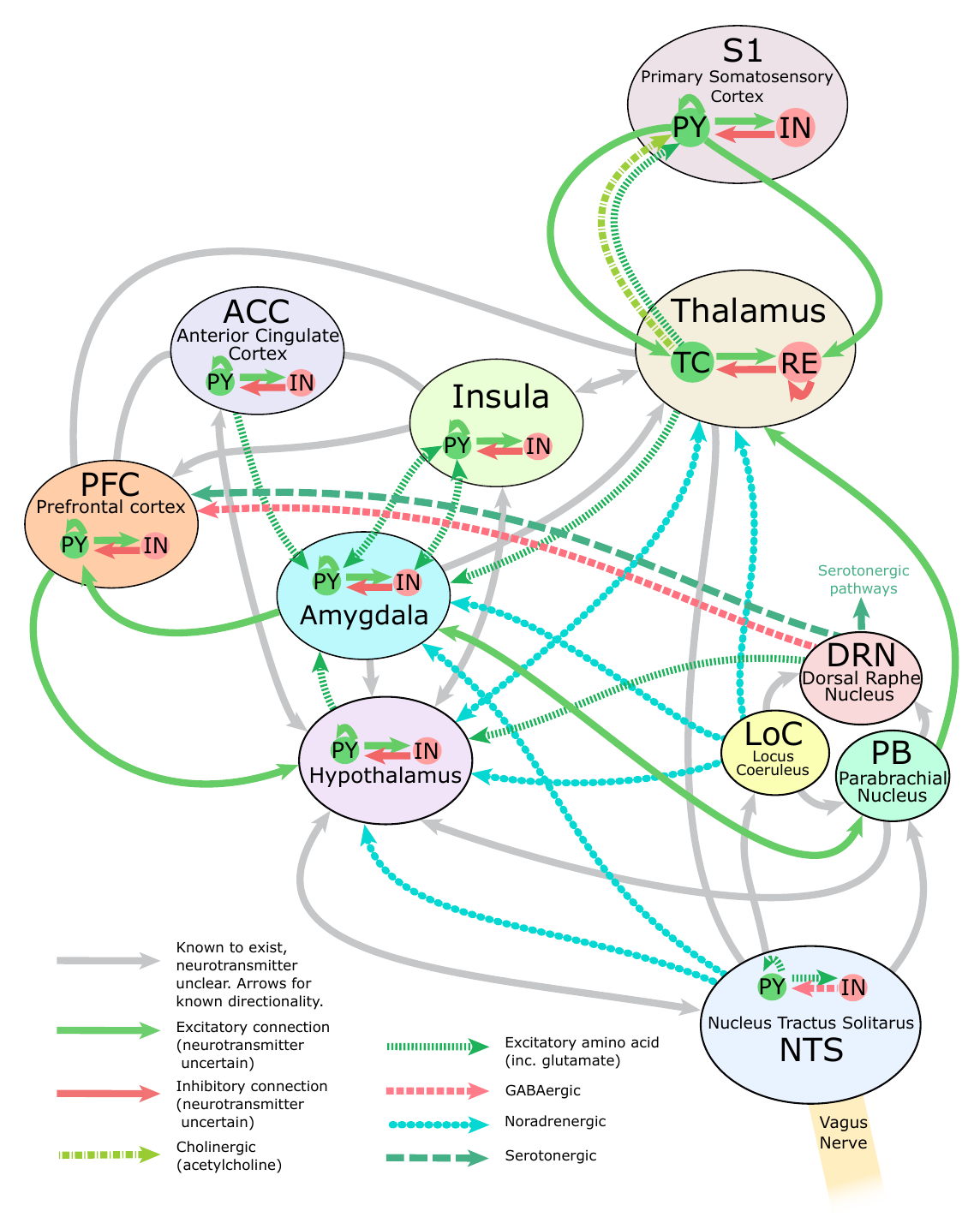}
	\caption{\textbf{Network of regions associated with VNS.} This figure shows brain regions associated with VNS and known connections between them. (Sources: NTS-LoC~\protect\citep{hachemVagusAfferentNetwork2018}, NTS-PB~\protect\citep{hachemVagusAfferentNetwork2018,herbertConnectionsParabrachialNucleus1990}, NTS-DRN~\protect\cite{hachemVagusAfferentNetwork2018}, NTS-Thalamus~\protect\citep{walkerRegulationLimbicMotor1999,ruffoliChemicalNeuroanatomyVagus2011}, NTS-Amygdala~\protect\citep{hachemVagusAfferentNetwork2018,GonzalezArancibia2019,maharjanImprovementOlfactoryFunction2018}, NTS-Hypothalamus~\protect\citep{GonzalezArancibia2019,maharjanImprovementOlfactoryFunction2018,jeanNoyauFaisceauSolitaire1991,gautronNeurobiologyInflammationassociatedAnorexia2009}, LoC-PB~\protect\citep{hachemVagusAfferentNetwork2018}, LoC-DRN~\protect\citep{hachemVagusAfferentNetwork2018}, LoC-Thalamus~\protect\citep{hachemVagusAfferentNetwork2018}, LoC-Amygdala~\protect\citep{hachemVagusAfferentNetwork2018,maharjanImprovementOlfactoryFunction2018}, LoC-Hypothalamus~\protect\citep{maharjanImprovementOlfactoryFunction2018,lemaireWhiteMatterConnectivity2011}, DRN-PB~\protect\citep{pritchardProjectionsParabrachialNucleus2000}, DRN-PFC~\protect\citep{bangForebrainGABAergicProjections2012}, PB-Hypothalamus~\protect\citep{pritchardProjectionsParabrachialNucleus2000,verberneNeuralPathwaysThat2014}, PB-Thalamus~\protect\citep{hachemVagusAfferentNetwork2018,pritchardProjectionsParabrachialNucleus2000,woolseyBrainAtlasVisual2017}, PB-Amygdala~\protect\cite{pritchardProjectionsParabrachialNucleus2000}, Thalamus-Insula~\protect\citep{woolseyBrainAtlasVisual2017,choNucleusAccumbensThalamus2013,mufsonThalamicConnectionsInsula1984,shiCorticalThalamicAmygdaloid1998}, Thalamus-Amygdala~\protect\citep{fischerLabellingAmygdalopetalAmygdalofugal1982}, Thalamus-S1~\protect\citep{taylorComputationalStudyStimulus2014,woolseyBrainAtlasVisual2017,barthSexHormonesAffect2015,middletonRevisedNeuroanatomyFrontalSubcortical2001}, Thalamus-Hypothalamus~\protect\citep{lemaireWhiteMatterConnectivity2011}, Thalamus-PFC~\protect\citep{robertsForebrainConnectivityPrefrontal2007,shiCorticalThalamicAmygdaloid1998,hachemVagusAfferentNetwork2018,mufsonThalamicConnectionsInsula1984,woolseyBrainAtlasVisual2017}, Insula-Amygdala~\protect\citep{Amaral1992,smithIntraAmygdaloidProjections1994,oharaDopaminergicInputGABAergic2003,sunEvidenceGABAergicInterface1994}, Insula-ACC~\protect\citep{hachemVagusAfferentNetwork2018,lichensteinAdolescentBrainDevelopment2016}, Insula-Hypothalamus~\protect\citep{lemaireWhiteMatterConnectivity2011,ongurPrefrontalCorticalProjections1998,woolseyBrainAtlasVisual2017}, Amygdala-PFC~\protect\citep{robertsForebrainConnectivityPrefrontal2007,reppucciOrganizationConnectionsAmygdala2016}, ACC-PFC~\protect\citep{hachemVagusAfferentNetwork2018,yanFunctionalAnatomicalConnectivity2012}, ACC-Hypothalamus~\protect\citep{ongurPrefrontalCorticalProjections1998}, PFC-Hypothalamus~\protect\citep{hachemVagusAfferentNetwork2018,lemaireWhiteMatterConnectivity2011,ongurPrefrontalCorticalProjections1998})
	}
	\label{Fig.1}
\end{figure}

We based our model on previously published work, which was able to replicate epileptiform spike and wave discharge (SWD) events that occur and terminate spontaneously with background noise~\citep{taylorComputationalStudyStimulus2014}. This model includes equations representing thalamic and cortical areas, and was used to make predictions about single pulse cortical stimulation as a means to elicit or disrupt seizure-like activity. In this paper we first explore the literature surrounding the brain regions thought to be implicated in VNS. We then use this information to extend the thalamocortical model to include this wider network of VNS-implicated regions, and allow us to capture the simulated impact of VNS on the brain. The network is visualised in Fig.~\ref{Fig.1}, which includes references to specific connections identified~\citep{hachemVagusAfferentNetwork2018, herbertConnectionsParabrachialNucleus1990,walkerRegulationLimbicMotor1999,ruffoliChemicalNeuroanatomyVagus2011,GonzalezArancibia2019,maharjanImprovementOlfactoryFunction2018,jeanNoyauFaisceauSolitaire1991,pritchardProjectionsParabrachialNucleus2000,bangForebrainGABAergicProjections2012,verberneNeuralPathwaysThat2014,woolseyBrainAtlasVisual2017,choNucleusAccumbensThalamus2013,shiCorticalThalamicAmygdaloid1998,fischerLabellingAmygdalopetalAmygdalofugal1982,barthSexHormonesAffect2015,middletonRevisedNeuroanatomyFrontalSubcortical2001,robertsForebrainConnectivityPrefrontal2007,Amaral1992,smithIntraAmygdaloidProjections1994,ongurPrefrontalCorticalProjections1998,lichensteinAdolescentBrainDevelopment2016}.

\subsection{Constructing a VNS network from literature}

\subsubsection*{Nucleus Tractus Solitarius}
The Nucleus Tractus Solitarius (also known as the Nucleus of the Solitary Tract or the Solitary Nucleus, abbreviated here as NTS) is a brainstem nucleus which receives direct input from the vagus nerve, and is therefore the first brain area to be impacted by VNS~\cite{zhangExcitatoryAminoacidReceptors1995,sessleExcitatoryInhibitoryInputs1973}. The NTS projects to other brainstem nuclei including the Locus Coeruleus and Parabrachial Nucleus which in particular are associated with VNS~\cite{hachemVagusAfferentNetwork2018}. Long range connections to the hypothalamus~\cite{jeanNoyauFaisceauSolitaire1991} and nuclei of the thalamus and amygdala~\cite{fanResearchProgressVagus2019} have also been described. 

\subsubsection*{Locus Coeruleus}
The Locus Coeruleus (LoC) is a region which has been associated with VNS effects in several studies: lesion experiments in rodents showed that damaging the LoC prevented the anti-convulsant effects of VNS~\cite{krahlLocusCoeruleusLesions1998} and activation of the LoC has been associated with VNS responders~\cite{detaeyeP3EventRelatedPotential2014}. The LoC connects proximally to the NTS and is the primary source of noradrenergic neurons in the brain, making it a region with influence on the wider network~\cite{hachemVagusAfferentNetwork2018}. 

\subsubsection*{Dorsal Raphe Nucleus}
The Dorsal Raphe Nucleus (DRN) is included in our model on a similar basis to the LoC, as it is a region with known close connections to the LoC, which is in turn connected to the NTS, and is a major source of serotonergic innervation of the wider brain~\cite{hachemVagusAfferentNetwork2018}, making it capable of widespread influence. Serotonin has been implicated in seizure thresholds~\cite{fanselowCentralMechanismsCranial2012} and VNS has been suggested as a treatment for depression where serotonin is again known to have a strong influence~\cite{carrRoleSerotoninReceptor2011,maceSelectiveSerotoninReuptake2000,cowenSerotoninDepressionPathophysiological2008}. We therefore include in the model this key region for the regulation of the serotonergic network. Possible future expansions of this model could better capture the DRN influence on this wider network beyond the regions modelled in our current system. 

\subsubsection*{Parabrachial Nucleus}
The Parabrachial Nucleus (PB) is another brainstem region with a known close connection from the NTS~\cite{herbertConnectionsParabrachialNucleus1990}, which connects closely with the LoC and DRN. It is also known to have diffuse outputs to the thalamus, amygdala and hypothalamus~\cite{hachemVagusAfferentNetwork2018} which are regions we have also included in this network.

\subsubsection*{Hypothalamus}
The hypothalamus is directly innervated by the NTS~\cite{fanResearchProgressVagus2019,GonzalezArancibia2019,hachemVagusAfferentNetwork2018,jeanNoyauFaisceauSolitaire1991} and connects with all other regions of interest in our model~\cite{ongurPrefrontalCorticalProjections1998,verberneNeuralPathwaysThat2014,pritchardProjectionsParabrachialNucleus2000,lemaireWhiteMatterConnectivity2011,hiokiVesicularGlutamateTransporter2010,maharjanImprovementOlfactoryFunction2018}. Another reason to include the hypothalamus is that there is evidence that VNS has an immunomodulatory effect via the hypothalamic-pituitary-adrenal axis~\cite{deherdtIncreasedRatSerum2009,okeaneChangesHypothalamicpituitaryadrenalAxis2005,gilElectricalChronicVagus2013}. Inflammation has been proposed to play a role in seizure generation~\cite{vezzaniRoleInflammationEpilepsy2011} and inflammatory markers may predict VNS response~\cite{majoieVagusNerveStimulation2011}. Other studies investigated the efficacy of VNS for epilepsy associated with hypothalamic hamartomas ~\cite{frostVagusNerveStimulator2000,murphyLeftVagalNerve2000}. 

\subsubsection*{Amygdala}

The amygdala is often associated with seizure activity, particularly in temporal lobe epilepsy~\cite{kullmannWhatWrongAmygdala2011}. The amgydala receives connections from the NTS directly as well as from the PB~\cite{pritchardProjectionsParabrachialNucleus2000} and noradrenergic innervation from the LoC~\cite{maharjanImprovementOlfactoryFunction2018,hachemVagusAfferentNetwork2018}. It is known to have connections with the hypothalamus~\cite{ongurPrefrontalCorticalProjections1998,lemaireWhiteMatterConnectivity2011}, insula~\cite{woolseyBrainAtlasVisual2017,hachemVagusAfferentNetwork2018,oharaDopaminergicInputGABAergic2003}, prefrontal cortex~\cite{reppucciOrganizationConnectionsAmygdala2016}, anterior cingulate cortex~\cite{lichensteinAdolescentBrainDevelopment2016,marusakYouSayPrefrontal2016} and thalamus~\cite{woolseyBrainAtlasVisual2017,fischerLabellingAmygdalopetalAmygdalofugal1982}, all of which are included in our model. 

\subsubsection*{Prefrontal Cortex}

The prefrontal cortex (PFC) is connected with the amygdala~\cite{reppucciOrganizationConnectionsAmygdala2016}, insula~\cite{oharaDopaminergicInputGABAergic2003}, hypothalamus~\cite{ongurPrefrontalCorticalProjections1998,Amaral1992,lemaireWhiteMatterConnectivity2011} and the anterior cingulate cortex~\cite{yanFunctionalAnatomicalConnectivity2012,hachemVagusAfferentNetwork2018}, as well as directly to the thalamus~\cite{robertsForebrainConnectivityPrefrontal2007}. An fMRI study of patients receiving VNS showed activation of the frontal cortex~\cite{liuBOLDFMRIActivation2003}, and decreased functional connectivity has been found between the prefrontal cortex and the cingulate cortex in VNS responders~\cite{Bodin2015,bartolomeiHowDoesVagal2016}. 

\subsubsection*{Anterior Cingulate Cortex}

Cingulate cortex has been implicated in a number of connectivity studies investigating VNS response~\cite{Bodin2015,barthSexHormonesAffect2015,workewychBiomarkersSeizureResponse2020}, and was shown to be activated in fMRI analysis of VNS recipients being treated for epilepsy~\cite{liuBOLDFMRIActivation2003}. Additionally, the anterior cingulate cortex connects to the insula~\cite{hachemVagusAfferentNetwork2018,lichensteinAdolescentBrainDevelopment2016}, PFC~\cite{yanFunctionalAnatomicalConnectivity2012,hachemVagusAfferentNetwork2018}, amygdala~\cite{marusakYouSayPrefrontal2016,Amaral1992,lichensteinAdolescentBrainDevelopment2016} and hypothalamus~\cite{ongurPrefrontalCorticalProjections1998}. 

\subsubsection*{Insula}

Connections between the insula and thalamic and temporal regions, as well as between the insula and brainstem and cingulate cortex, have been implicated in VNS response~\cite{workewychBiomarkersSeizureResponse2020,zhuFunctionalConnectivityStudy2020}. The insula is interconnected with other regions of interest in the model: the PFC~\cite{oharaDopaminergicInputGABAergic2003}, the Anterior Cingulate Cortex~\cite{hachemVagusAfferentNetwork2018,lichensteinAdolescentBrainDevelopment2016}, the amygdala~\cite{smithIntraAmygdaloidProjections1994, Amaral1992,woolseyBrainAtlasVisual2017,hachemVagusAfferentNetwork2018}, hypothalamus~\cite{ongurPrefrontalCorticalProjections1998}, and the thalamus~\cite{choNucleusAccumbensThalamus2013,mufsonThalamicConnectionsInsula1984,shiCorticalThalamicAmygdaloid1998}. The insula has also been shown to be activated by VNS in fMRI~\cite{liuBOLDFMRIActivation2003}.

\subsection{Determining relative connection strength from neuroimaging data}

In order to determine connection weights in the model, we utilise diffusion tensor imaging data from the human connectome project. We computed connectivity networks using DSI Studio~\citep{yehNTU90HighAngular2011,yehPopulationbasedTracttoregionConnectome2022} and the HCP842 tractography atlas overlaid with fractional anisotropy (FA) maps, sampling the FA along each tract. The Desikan-Killiany parcellation~\citep{desikanAutomatedLabelingSystem2006} was used as grey matter regions of interest and we considered two regions connected if a tract ended in both regions. The average connectivity across the controls was computed to form a representative value in this study. However, in future studies we anticipate that these values could be used from individual patient data. The control averaged values were then scaled relative to the model connection weight between S1 and the Thalamus, as determined in~\cite{taylorComputationalStudyStimulus2014}, such that this weight was equal to the value in the original model, and all other weights take on values relative to it. All connectivity values are included in Table S0.1 of the supplementary materials section. 

\subsection{Building the computational model}

The basis for our model, is a four-population NMM representing excitatory ($PY$) and inhibitory ($IN$) neocortical populations, specifically labelled here as primary somatosensory cortex (S1), as well as the excitatory thalamocortical nucleus ($TC$) and inhibitory reticular ($RE$) nucleus of the thalamus~\citep{taylorComputationalStudyStimulus2014}. These, in turn, are modelled using a development of the Amari neural field equations~\citep{amariDynamicsPatternFormation1977,taylorSpatiallyExtendedModel2011,taylorModelStimulusInduced2013}. The deterministic version of this model is described by the following differential equations:

\begin{equation}
\begin{aligned}
    \frac{\mathrm{d}PY}{\mathrm{d}t} \;&=\; \uptau_1(h_{PY}-PY+C_1f[PY]-C_3f[IN]+C_9f[TC])\\[10pt]
    \frac{\mathrm{d}IN}{\mathrm{d}t} \;&=\; \uptau_2(h_{IN}-IN+C_2f[PY])\\[10pt]
    \frac{\mathrm{d}TC}{\mathrm{d}t} \;&=\; \uptau_3(h_{TC}-TC+C_7f[PY])-C_6s[RE]\\[10pt]
    \frac{\mathrm{d}RE}{\mathrm{d}t} \;&=\; \uptau_4(h_{RE}-RE+C_8f[PY])-C_4s[RE]+C_5s[TC],\\[10pt]   
\end{aligned}
\end{equation}
where: $h_{PY},\:h_{IN},\:h_{TC},\:h_{RE}$ are input parameters; $\uptau_{1},\;\ldots, \uptau_{4}$ are time constants; $C_{1},\:\ldots,\:C_{9}$ are connection weights between populations; and $f$ and $s$ are the activation functions:
\begin{align}
    f[u]&=\frac{1}{1+\varepsilon^{-u}},\\[10pt]
    s[u]&=au+b,
\end{align}
for $u=PY,IN,TC,RE$. The steepness of the sigmoid function, $s$, is determined by $\varepsilon$. The mean activity of the $PY$ and $IN$ populations are taken to represent the local field potential (LFP) of the cortex, or an electroencephalogram (EEG) channel measured from the local cortical surface or overlying scalp. Gaussian noise can be added to the $TC$ population input in order to induce spontaneous seizure-like episodes in the system, as in the original model~\citep{taylorComputationalStudyStimulus2014}.

From this starting point, we modelled the additional brain regions, identified in section 3.1, as pairs of excitatory and inhibitory populations, analogous to those of S1. The connection weights determined in section 3.2 are applied to the connections identified between the populations. Except for where specific evidence exists to the contrary, projections between areas are assumed to be between the excitatory populations.  Where the directionality of a tract is not known, half of the weight is assigned to either direction. Thresholding is applied such that exceptionally weak connection weights are ignored.

We first simulated the model deterministically, without the addition of random noise, in order to assess its bifurcation structure over a range of parameters, using the MATLAB ODE45 adaptive time step solver in its default settings. Parameter sweeps were performed with ascending and descending values starting from random points in the state space, and the local minima and maxima plotted. 

For the stochastic versions of the model, we choose to retain the $TC$ population as the site of noise injection as, not only is this in keeping with previous literature~\citep{robinsonDynamicsLargescaleBrain2002,breakspearUnifyingExplanationPrimary2006,martenDerivationAnalysisOrdinary2009}, but also has less direct effect on the measurement at S1, allowing easier detection of seizures, while still being able to influence the core dynamics of the model. Noise is normally distributed with a mean of zero, while the standard deviation was adjusted to 0.72, which produced clinically plausible seizure duration and frequency.

A solver utilising the Euler-Maruyama method was used to evaluate stochastic versions of the model. Its fixed time step was determined through comparing zero noise simulations against those from the deterministic system's variable step solver, and by ensuring consistency between stochastic behaviour and the bifurcation behaviour of the deterministic system. A step size of 100~\textmu s was the minimum required to give consistent results, and was therefore used throughout.

For stochastic simulations we generated, for each set of parameters, a continuous time series of 10,000~s (approximately 27.8~hours) of simulated time using identical noise sequences in each case. For consistency, the noise was generated by concatenating consistent epochs of 100~s duration, with the random number generator (RNG)  seeded at 1 for the first epoch, incremented for each subsequent epoch. Specifying the RNG generation algorithm allowed consistency between versions of MATLAB running on desktop and HPC cluster. Each simulation started at the FP and was allowed to evolve thereafter for the duration of the simulation. 

\subsection{Seizure detection}
The FP was determined at the default parameters and was found to change very little in the parameter ranges scanned. This was also the case for the LoC where present. The same FP values were therefore used throughout. For all time series, the euclidean distance (ED) of the S1 populations from the FP is calculated. We considered other combinations of the 22 neuronal populations to contribute to this measure, but the S1 populations give the clearest signal. The ED time series was smoothed by applying a moving mean with a 2-second window. The system is determined to be in a state of seizure if the smoothed ED is greater than a value of 0.15. Both the threshold and the duration of the moving mean window were determined empirically to minimise rapid switching between the normal and seizure states where possible. The seizure detection method is summarised in Figure S0.1 of the supplementary materials.

Using the same parameters used in the deterministic version of the model, the noise was scaled to the point where the system spent the vast majority of time in the non-seizure state, with a clinically plausible seizure frequency and duration. However, having infrequent seizures necessitates longer simulations to obtain an interpretable quantity of events, so a compromise was made to simulate the more severe end of the clinical spectrum.

\subsection{Simulating vagus nerve stimulation}
Having established a value of NTS\textsubscript{PY} placing the system within the seizure prone region, simulated VNS is applied. This was modelled as a 30 Hz square wave with a pulse width of 500~\textmu s, consistent with clinical practice, and added to the input of NTS\textsubscript{PY}. Repeated runs were performed with increasing amplitudes of VNS, starting at zero.

Full details of all parameters are given in Tables S0.1 and S0.2 of the supplementary materials. Code to reproduce the findings is available at: https://github.com/johningham/VNS\_model\_code.

\section{Results}

\subsection{The deterministic system}
When simulating without the addition of noise, our model produced a very similar behaviour to the original model on which it is based. With our chosen set of parameters, the system demonstrated two distinct stable states, a stable FP and a period-2 LC with a biphasically oscillating time series resembling SWD (Figure 2a,b). All 22 populations remained strongly coupled. Time series and phase plots for these states are shown in Figure 2. 

Figure 2d also shows the effect of varying the background input of the excitatory population of NTS\textsubscript{PY}, which receives afferent connections from the vagus nerve \textit{in vivo}. At lower values of NTS\textsubscript{PY} input, the system is bistable between the FP and LC while, at higher values, it is monostable at the FP only. It is interesting to note that the location of the FP and the trajectory of the LC, where it exists, do not substantially change with NTS\textsubscript{PY}. 

\subsection{The stochastic system without VNS}
 Noise was applied such that the system spent a maximum of 0.309\% of the total time in the seizure state, as measured over 27.8 hours of simulated time. This was spread over 14 distinct seizures with a mean duration of 61.3s, consistent with the more severe end of the clinical range. We performed repeated simulations, varying the value of NTS\textsubscript{PY} input, using identical noise for each simulation. With a sufficiently high value of NTS\textsubscript{PY} input ($>$0-0.52), the system remained monostable in the normal state, as with the deterministic version of the model. Conversely, at lower values ($<$-0.65) the system can reach either state. In the stochastic case, there is an intermediate reduction in the proportion of time spent in seizure before all seizure activity disappears. This is shown in Figure~3a, together with an example of a time series of the simulated LFP as the system transitions from the normal state into seizure (Figure 3b). This particular example was achieved with an NTS\textsubscript{PY} input value of -0.7, which was the value chosen as the background value when VNS was applied.

\subsection{The stochastic system with VNS}
Increasing the amplitude of simulated VNS reduced the total proportion of time spent in seizure in a dose-dependent manner (Figure 4a). This is consistent with expectations from the clinical application of VNS in epilepsy. VNS causes some seizure events to terminate earlier than they otherwise would, also replicating the effect of VNS seen clinically (Figure 4b,c).

\section{Discussion}

In this study we developed a model of seizures in the human brain, incorporating areas of significance for VNS treatment. The model shows transitions between background, and seizure-like states, whilst being amenable to simulating stimulation. We successfully reproduced the clinical observation of reduced seizure duration, with stimulation of the vagus nerve \textit{in silico}.

The deterministic versions of our model behave in a very similar way to the original model, generating monostable (FP) and bistable (FP and LC) regions within the parameter space. The addition of noise produces a stochastic version of our system with very analogous behaviour. This can be tuned to represent a brain that is prone to seizures to various extents or one in which a seizure will not occur, again just as with the base model.

Computational brain modelling has a long history that parallels scientific understanding of normal and disordered brain function of humans and other species. It has been applied successfully at scales from ion channels across axonal membranes~\citep{hodgkinQuantitativeDescriptionMembrane1952}, through models of single neurons at various levels of abstraction~\citep{mccullochLogicalCalculusIdeas1943,izhikevichSimpleModelSpiking2003}, small scale neuronal circuits, neuronal populations~\citep{wilsonExcitatoryInhibitoryInteractions1972,wilsonMathematicalTheoryFunctional1973,jansenElectroencephalogramVisualEvoked1995}, cortical regions, and networks through to the whole brain, with each successful model abstracting the properties important at that scale~\citep{lyttonComputerModellingEpilepsy2008,depannemaeckerModelingSeizuresSingle2021}. Many general NMMs of the brain have been successfully applied to the field of epilepsy, such as the Wilson-Cowan~\citep{wilsonExcitatoryInhibitoryInteractions1972,wilsonMathematicalTheoryFunctional1973,wilsonEvolutionWilsonCowan2021a} and Jansen-Rit~\citep{jansenElectroencephalogramVisualEvoked1995,wendlingRelevanceNonlinearLumpedparameter2000}, while newer NMMs have been designed from the outset with the epilepsy in mind~\citep{wendlingEpilepticFastActivity2002,suffczynskiDynamicsNonconvulsiveEpileptic2004}, as have epilepsy specific models at other scales~\citep{traubSingleColumnThalamocorticalNetwork2005}. Connecting multiple instances of NMMs, or other models, has proven fruitful~\citep{wangMechanismsUnderlyingDifferent2017,depannemaeckerModelPropagationSeizure2022} to generate more complex behaviours and to model phenomena such as spread of seizure activity. While models such as those above are typically derived from established physiological principles, with parameters tuned to approximate the modelled phenomena, another approach is to use mathematical first principles to reproduce the physiologically known seizure dynamics, with subsequent identification of the model parameters with known biological correlates. The Epileptor~\citep{jirsaNatureSeizureDynamics2014} would be an example of this approach. 

The thalamocortical model~\citep{taylorComputationalStudyStimulus2014}, in particular, has been used: to investigate the dynamics created from coupled instances~\citep{fanRichDynamicsInduced2018}, including identifying a proposed regulatory mechanism~\citep{zhangRegulatoryMechanismAbsence2021}; as inspiration for an ensemble model of SWD behaviour in a genetic rodent model~\citep{medvedevaDynamicalMesoscaleModel2020a}; for analysis of EEG signals~\citep{medvedevaEstimatingComplexitySpikewave2020}; to create a model measuring proximity to seizure~\citep{deebaUnifiedDynamicsInterictal2019} to model brain stimulation in seizure-like states\citep{wangStimulationStrategiesAbsence2019,yanControlAnalysisElectrical2020}, including by closed-loop control techniques~\citep{geRobustClosedloopControl2019,zhouSeizureSuppressionThalamocortical2020,qianNovelNonsingularIntegral2022}; to form the basis for a proposed radio-technical model for a hierarchical neural network~\citep{egorovSimulationEpileptiformActivity2021}; to explore the potential dynamical role of electromagnetic induction in seizures~\citep{zhaoDynamicalRoleElectromagnetic2021}; and to simulate Glucose Transporter Deficiency induced epileptic seizures in a network of neurons ~\citep{leslieModelingSimulationNetwork2022}.

It is hoped that this model will contribute to the general understanding of the mechanism of VNS in epilepsy with the expectation that additional testing against experimental findings should result in further refinements. Moreover, a number of specific potential uses are apparent. 

A first potential application is to devise and test short term seizure control protocols in a responsive manner. In addition to the continuous scheduled stimulation cycle, many devices have a ``Magnet Mode" enabled which allows an instantaneous stimulation to be administered by a patient or carer to treat an impending or ongoing seizure~\citep{fisherVagusNerveStimulation2015}. Some recent devices also have an ``AutoStim" feature, which performs a similar function in response to heart rate changes characteristic of a seizure. The possibility that VNS may have multiple mechanisms of action working over different timescales~\citep{carronLatestViewsMechanisms2023}, suggests that it may be optimal to consider different stimulation parameters for each mode of operation, beyond the current practice of allowing marginally higher amplitude stimulation in magnet mode~\citep{fisherResponsiveVagusNerve2021}.

We also foresee an application to personalised models based on a patient's own brain imaging to constrain model parameters. It currently almost impossible to predict which individuals might respond best to VNS~\citep{piresdopradoPredictiveFactorsSuccessful2023}. This approach may be useful for prognostication, aiding in patient selection and preoperative counselling. We also envisage a role in guiding stimulation protocol choice for individuals. This is currently performed on a trial and error basis during scheduled clinic visits~\citep{thompsonReviewParameterSettings2021}. By having an individualised model on which multiple stimulation parameter changes could be optimised in a short period of time, it is hoped that potential options could be narrowed down before trying them clinically.

Individualised models have already been proposed in epilepsy management, especially in the field of surgery, where accurately selecting what tissue to resect is crucially important~\citep{goodfellowEstimationBrainNetwork2016,sinhaPredictingNeurosurgicalOutcomes2017}. In particular, multi-scale models~\cite{kuhlmannRoleMultiplescaleModelling2015}, where models of phenomena occurring at adjacent spatial or temporal scales are combined, have been developed which can have additional explanatory power than simpler models~\citep{wendlingEpilepticFastIntracerebral2003,wendlingInterictalIctalTransition2005}.

There are some limitations to our model. As with all models, it is, by necessity, a simplification with various assumptions. Our additional regions have each been modelled as two populations, sharing identical parameters with those used for S1. 
Many of the new regions, especially those which are subcortical or pathological, will have different cell types, architectures, neurotransmitters and physiological parameters, but have not been modelled as extensively as the neocortex.
It may be worth pursuing further refinements, for example by constraining regional parameters based on pathology evident from MRI or EEG~\citep{hutchingsPredictingSurgeryTargets2015,timotheeproixIndividualBrainStructure2018}.

The model only captures the effects of VNS over very short timescales (seconds) and ignores potential role of plasticity,  modulation of inflammatory processes, and other potential mechanisms~\cite{carronLatestViewsMechanisms2023} that might explain the effect of VNS on timescales up to months and years~\citep{giannakakisComputationalModellingLongterm2020}. Nevertheless, the evidence suggests that there remains a very definite benefit from stimulation in the immediate term, as demonstrated by the effectiveness of magnet mode~\citep{fisherVagusNerveStimulation2015}, and AutoStim~\citep{fisherAutomaticVagusNerve2016}. The model therefore retains its relevance, especially when applied to treatment over shorter time-scales.  

In the stochastic modelling, for both increasing background NTS\textsubscript{PY} input and simulated VNS, the effect on seizure frequency is not as clear as it is on total seizure duration. This appears to be due to long seizures being broken into multiple short seizures rather than new short seizures being produced at times where runs for the parameters with the lower values were normal. 

VNS is used in a group of patients who have been unable to obtain control of their seizures with medication and, in some cases, surgery or other types of stimulation. As such it is especially important that this treatment modality can be fully understood and used optimally. This model will serve as a starting point to investigate the mechanisms of VNS and practical benefits to patients. 

\section{Acknowledgements}
We thank members of the Computational Neurology, Neuroscience \& Psychiatry Lab (www.cnnp-lab.com) for discussions on the analysis and manuscript; This work was supported by the Engineering and Physical Sciences Research Council [grant number 2595464]. P.N.T. and Y.W. are both supported by UKRI Future Leaders Fellowships (MR/T04294X/1, MR/V026569/1). Sadegh Soudjani is supported by the following grants: EPSRC EP/V043676/1, EIC 101070802, and ERC 101089047. 

\newpage

\begin{figure}
	\centering
	\includegraphics[width=\textwidth]{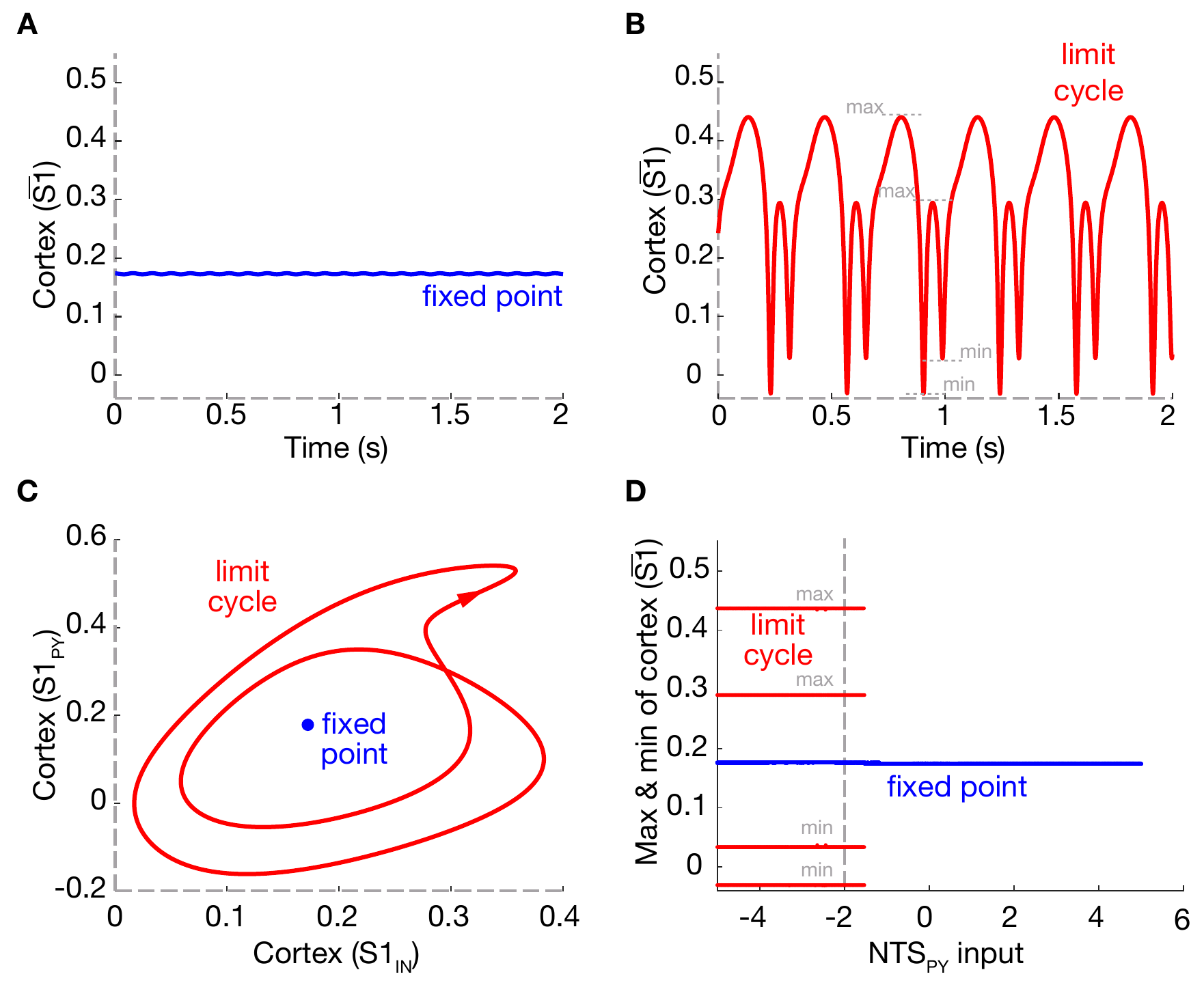}
	\caption{\textbf{Behaviour of the Deterministic System.} Simulated LFP time series at S1, generated from the mean activities of S1\textsubscript{PY} and S1\textsubscript{IN}, at both the FP (A), and LC (B).These are both states are shown in a phase plot (C). The bifurcation behaviour of the system as the value of NTS\textsubscript{PY} input varies is shown in (D).
	}
	\label{Fig.2}
\end{figure}

\begin{figure}
	\centering
	\includegraphics[width=\textwidth]{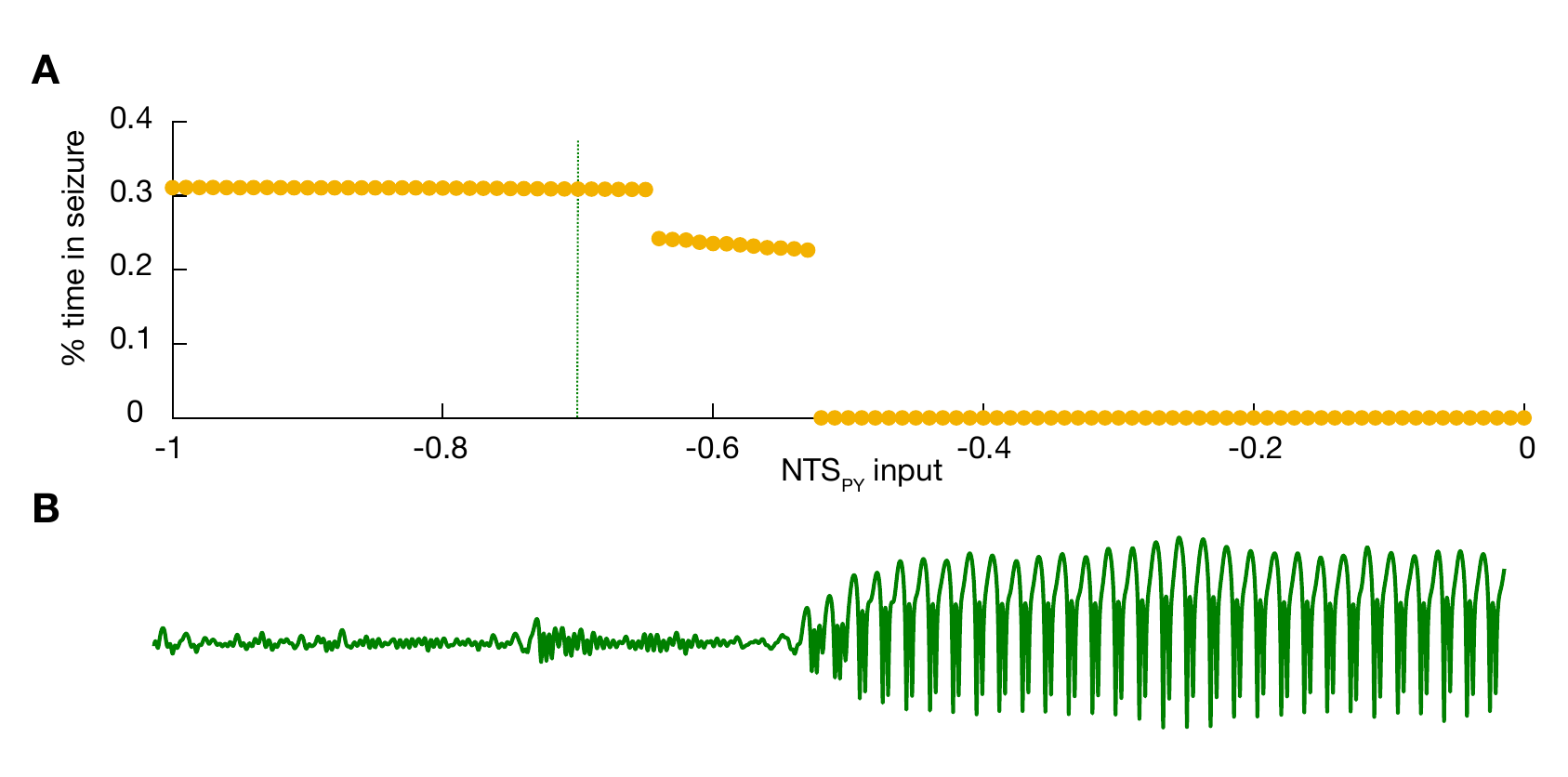}
	\caption{\textbf{The Stochastic System without VNS.} (A) The effect of varying the background input of NTS\textsubscript{PY} on the proportion of the system's seizure time over an extended run, and (B) a short sample of the simulated LFP in the S1 region at a NTS\textsubscript{PY} input value of -0.7, showing a transition from baseline to a seizure state. This value is used as as the background level in the model once VNS is applied. 
	}
	\label{Fig.3}
\end{figure}

\begin{figure}
	\centering
	\includegraphics[width=\textwidth]{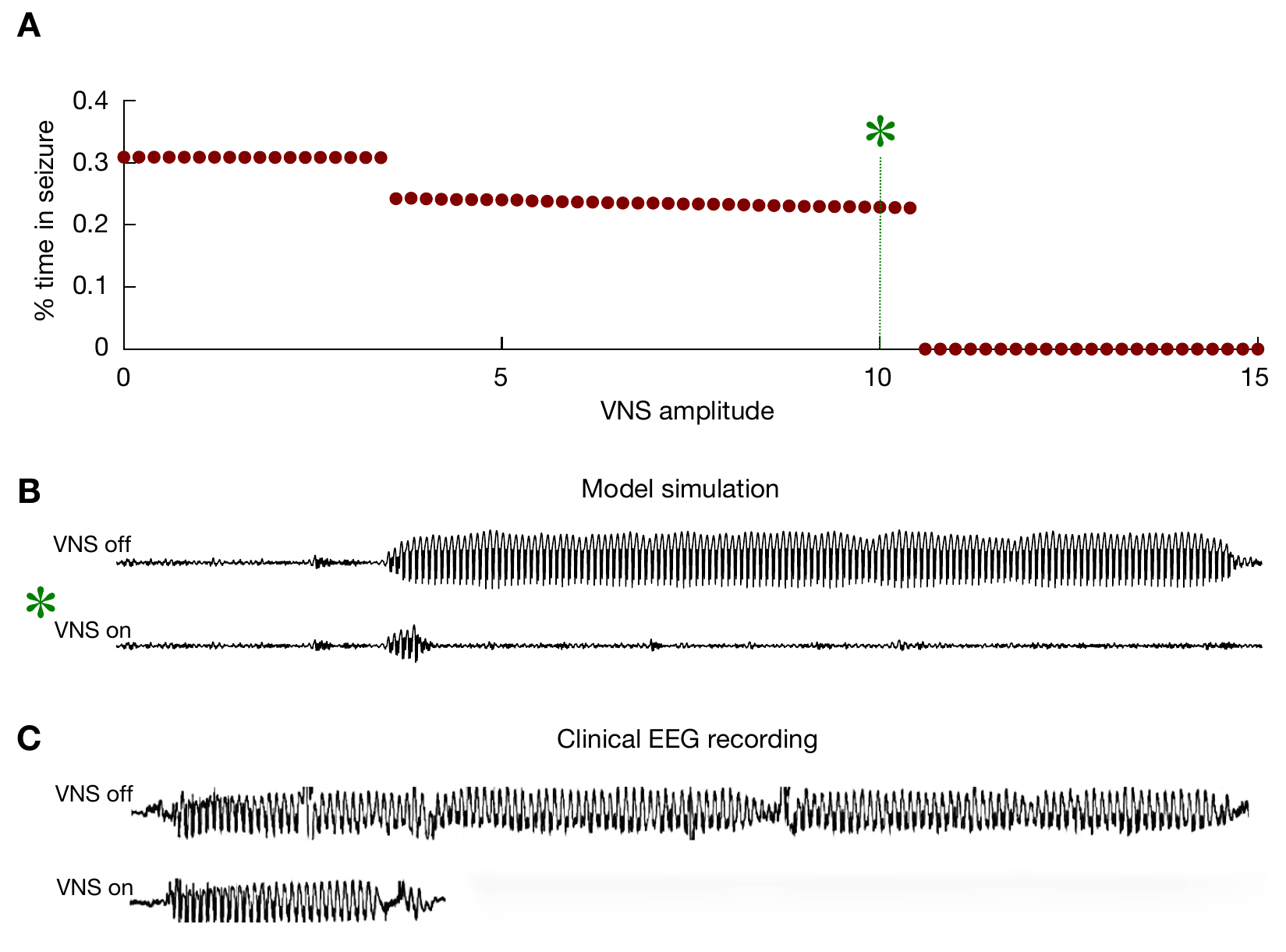}
	\caption{\textbf{The Application of VNS.} (A) The effect on the seizure time of applying various amplitudes of simulated VNS to the model. The time series plots in (B) show simulated LFP data, with the upper plot showing the system without VNS applied and the lower plot showing the effect of VNS, with an amplitude of 10, over the same time interval. The seizure initiates at the same point in either case, but is terminated much earlier in this instance when VNS is applied. This can be compared with with the plots in (C), which are patient EEG data with and without VNS applied. Panel (C) modified with permission from~\protect\cite{franzoniVNSDrugResistant2010}.
	}
	\label{Fig.4}
\end{figure}

\newpage
\bibliography{references}
\end{document}